\documentclass[aps,prd,preprintnumbers,nofootinbib,twocolumn]{revtex4}
\usepackage{bm}
\usepackage{latexsym}
\usepackage{dcolumn}
\usepackage{amsmath,amsfonts,amssymb}
\usepackage{graphicx,epsfig}
\usepackage{psfrag}
\usepackage{amsthm}

\def\be {\begin{equation}}
\def\ee {\end{equation}}
\def\bea {\begin{eqnarray}}
\def\eea {\end{eqnarray}}
\def\bc {\begin{center}}
\def\ec {\end{center}}
\def\bfg {\begin{figure}}
\def\efg {\end{figure}}
\def\bi {\begin{itemize}}
\def\ei {\end{itemize}}

\def\la {\label}
\def\le {\left}
\def\ri {\right}

\def\no {\noindent}

\def\vs {\vspace}

\def\a  {\alpha}

\def\beq{\begin{equation}}
\def\eeq{\end{equation}}
\def\br{\begin{eqnarray}}
\def\er{\end{eqnarray}}
\newcommand{\eel}[1] {\label{#1}\end{equation}}

\newcommand{\bdm}{\begin{displaymath}}
\newcommand{\edm}{\end{displaymath}}

\begin{document}

\title{
Bose-Einstein condensation as an alternative to inflation
}

\author{Saurya Das} \email[email: ]{saurya.das@uleth.ca}

\vs{0.3cm}

\affiliation{
Department of Physics and Astronomy,
University of Lethbridge, 4401 University Drive,
Lethbridge, Alberta, Canada T1K 3M4 \\
{}\\
{\bf This essay received an Honorable Mention in the 2015 Gravity Research Foundation Essay Competition}
}
%

\begin{abstract}
It was recently shown that gravitons with a very small mass
should have formed a Bose-Einstein condensate in the very early Universe,
whose density and quantum potential can account for the dark matter and dark energy in
the Universe respectively.
Here we show that the condensation can also naturally explain the observed large scale homogeneity and isotropy
of the Universe.
Furthermore gravitons continue to fall into their ground state within the condensate at every epoch,
accounting for the observed flatness of space at cosmological distances scales.
Finally, we argue that the density perturbations due to
quantum fluctuations within the condensate give rise to a scale invariant spectrum.
This therefore provides a viable alternative to inflation, which is not associated with
the well-known problems associated with the latter.
%
%
\end{abstract}

\maketitle

It is believed that our Universe started around the time of the so-called
the Big Bang (BB) singularity, continued to expand,
and is currently doing so at an accelerating rate,
driven by a small cosmological constant or Dark Energy (DE) \cite{perlmutter,riess,wmap,bao}.
The latter constitutes about $70\%$ of all matter/energy, while the rest is shared by
Dark Matter (DM), about $25\%$ and visible matter, about $5\%$.
The observed incredible degree of homogeneity and isotropy
(`horizon problem') and spatial flatness (`flatness problem') of our Universe
are normally attributed to inflation, a proposed short but rapid phase of exponential expansion
soon after the BB, which smoothed out all unevenness \cite{guth,linde}.
Inflation also appears to give a ready explanation for the lack of observed GUT monopoles in the universe.
However, a number of serious problems associated with inflation
have been pointed out, including
the unknown origin and nature of a scalar field and the class of potentials
required for the process, which appears improbable, although not impossible, requiring
an enormous fine-tuning, and the ill-understood
energy transfer mechanism to the current contents of the Universe
\cite{earman,penrose,turok,gibbons,branden1,steinhardt,steinhardt2,steinhardt3}.
Furthermore recent analysis of observed data do not seem to support the simplest inflationary models
\cite{planckbicep}.
Thus a number of alternative theories have been proposed
\cite{ekpy,stringgas,mukhanov}.
Following some earlier work by the current author and collaborators \cite{sd,dasessay,alidas,dasbhaduri},
in this article we propose a much simpler explanation
of the above puzzles, one that requires few speculations and almost no fine tuning.
%
%
We argue that quanta of gravity, or gravitons with a tiny mass (but consistent with observations and theory)
form a Bose-Einstein condensate (BEC) in our Universe in the earliest epochs.
This is described by a single {\it macroscopic} quantum ground state of the size of the universe,
which as shown in \cite{dasbhaduri} can correctly account for both DM and DE.
Further by its very nature, it is homogeneous and isotropic, a
property which persists as the Universe expands to its current epoch and beyond.
In addition, since its constituent gravitons are all in their ground states with little or no momentum,
they do not propagate as force carriers. This manifests itself as zero spatial curvature at any epoch,
just as observed.
We also argue that the density perturbation spectrum is scale-invariant in this model.
Finally, without the need for additional fields, the problem of energy transfer to current contents
is absent.

We first review some essential results from \cite{sd,dasessay,alidas,dasbhaduri}.
Starting with the quantum corrected Raychaudhuri equation
obtained in \cite{sd} by assuming a fluid or condensate filling our Universe,
and described by a wavefunction
$\phi={\cal R}e^{iS}$ (${\cal R} (x^\a),S (x^a)=$ real functions),
the quantal (Bohmian) trajectories, defined by the
velocity field $u_a=\hbar \partial_a S/m$,
replace the classical geodesics \cite{bohm},
and it was found that the quantum corrected second order Friedmann equation
for the scale factor $a(t)$ is given by
($h_{ab} \equiv g_{ab} - u_a u_b$)
\cite{dasessay,alidas}
\bea
\frac{\ddot a}{a} =&& - \frac{4\pi G}{3} \le( \rho + 3p \ri) +
\frac{\Lambda c^2}{3} \nonumber \\
 && + \frac{\hbar^2}{3 m^2} h^{ab} \le( \frac{\Box {\cal R}}{\cal R} \ri)_{;a;b}
+\frac{\epsilon_1\hbar^2}{m^2}h^{ab}R_{;a;b}
~.
\la{frw1}
\eea
In \cite{dasessay,alidas, dasbhaduri} it was postulated that the condensate is nothing but a BEC of
gravitons of a tiny mass
\footnote{in \cite{dasbhaduri} axions were also explored.
Axions would still solve the homogeneity problem but not the flatness problem.
We ignore the possibility of axions here.}.
In fact it was shown in \cite{dasbhaduri} that the critical temperature
of such a condensate is given by
\bea
T_c = \frac{3}{m^{1/3} a}~K
\la{tc1}
\eea
such that for $m \lesssim 1~eV$, the critical temperature exceeds the temperature of the Universe, given by
$T=2.7/a~K, \forall a$,
meaning that most gravitons would have dropped to their ground state, and
a BEC would have formed in the very early universe.
Further while the first ${\cal O}(\hbar^2)$ term in Eq.(\ref{frw1}),
$\Lambda_Q \equiv  \frac{\hbar^2}{m^2 c^2} h^{ab} \le( \frac{\Box {\cal R}}{\cal R} \ri)_{;a;b}$
can be interpreted as the quantum mechanical contribution to the cosmological constant,
giving a correct estimate of the observed value, if $m=10^{-32}~eV$
(for which $T_c = 10^{11}/a \gg T,~ \forall a$),
the second ${\cal O}(\hbar^2)$ term implied an infinite age of our Universe, stretching
far beyond the time of the BB.
On the one hand, the above tiny graviton mass is consistent with theory and observations
\cite{zwicky,mukhanov2,oda,derham,hassan,mann,majid,hinterbichler,graviton}
while on the other, the above model is in agreement with all predictions of the
standard BB theory from the moment the Universe was of Planck size
\footnote{
For earlier work on BEC and superfluids in cosmology, see
\cite{sudarshan,hu,morikawa,moffat,wang,boehmer,sikivie,velten,dvali,chavanis,kain,suarez,ebadi,laszlo1,bettoni,gielen,schive,davidson}.
}.

\vs{0.2cm}
\noindent
{\bf Horizon problem}

Eq.(\ref{tc1}) shows that $T_c(a) \gg T(a)~\forall a$,
and that a BEC spreading across the Universe would have formed
almost right after the BB, if there was indeed a BB
(when both $T_c, T \rightarrow \infty$, but the former still far exceeds the latter).
On the other hand, if there was no BB, the scale factor was very small but never zero,
and the age of our Universe is infinite,
a possibility suggested in \cite{alidas}, then the BEC would have formed much earlier.
%
%
Since BEC has infinite heat conductivity, total
thermal equilibrium and temperature equalization
would have been established almost simultaneously throughout the (tiny) Universe.
Causality is never violated, the state of the Universe before BEC formation no longer matters,
and the BEC can be considered as the initial state for all practical purposes,
which evolves eventually to our current Universe.
This was also noted earlier in \cite{sudarshan}.
Also as noted in \cite{dasessay,alidas,dasbhaduri}, the resulting highly homogeneous and isotropic
condensate can be described by a macroscopic wavefunction, e.g. suitably approximated by
a shallow Gaussian \cite{brack} or having a shallow parabolic profile \cite{rogel}.
Furthermore, also as noted in \cite{dasbhaduri},
as the temperature of the Universe falls, more and more gravitons
fall in the condensate following the relation $N_B/N = 1 - (T/T_C)^3$, where
$N_B$ and $N$ are the number of gravitons in the condensate and the total number of
gravitons (inside $+$ outside the condensate) respectively.
$T_C \gg T$ implies $N \approx N_B$, i.e. there are hardly any gravitons left
outside the condensate at cosmological scales.
The condensate continues to grow, maintaining homogeneity and isotropy at every instant.
In other words, no fine-tuned initial conditions for our observable
Universe are required.
The situation is depicted in Figure 1.
The past light cones of CMBR photons at recombination
(within the bigger past light cone from a point $O$
in the current epoch) separated by more than a couple of degrees do not overlap in the
standard BB model without inflation (Figure 1(a)); therefore the observed
homogeneity, isotropy and uniformity of CMBR temperature accurate to $1$ part
in about $10^5$ is unexplained.
Figure 1(b) shows an extension of the bigger past light cone by virtue of
an exponential expansion, which now brings the smaller past light cones
in causal contact, thereby explaining the above.
Figure 1(c) shows how this can also be achieved if a BEC forms early on (black patch)
as explained above.
Figure 1(d) depicts if there was no BB or initial singularity,
in which the very large, or probable infinite age of the Universe gives ample opportunity
for the BEC to form and temperature to equalize throughout.
\null
\begin{center}
\begin{figure}
\includegraphics[scale=0.46,angle=0]{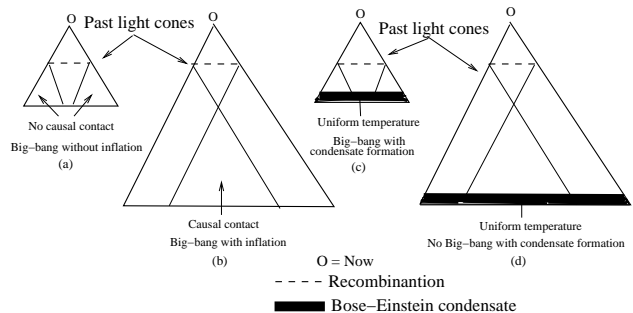}
\caption{(a) Big Bang without inflation,
(b) Big Bang with inflation,
(c) Big Bang with condensate formation, and
(d) No Big Bang with condensate formation.
Conformal time along the vertical.}
\label{solution}
\end{figure}
\end{center}

\noindent
{\bf Flatness problem}

In a quantum theory of gravitation, gravitons are the force carriers, just as photons are for
the electromagnetic force in quantum electrodynamics.
Indeed, as was shown e.g. in \cite{feynman} and \cite{deser1}, curvature of spacetime (hence
gravitational force) can be entirely built up by graviton fluctuations and their consistent
self-couplings, such that they contribute to the full divergence-free energy-momentum tensor.
Then since as argued above,
the overwhelming majority of gravitons are `frozen' in their ground states
in the BEC with little or no momenta (they only have rest energies,
are cold and can be identified with DM),
they do not propagate, there is no gravity nor gravity waves, and space is
flat at any given instant of time.
This can also be expressed in terms of perturbations $h_{ij}$ over the flat metric (no gravity),
or `ripples in spacetime' which are absent at large scales, and the
$3$-dimensional Riemann tensor
\bea
R^{(3)}_{ijkl}=\frac{1}{2}\le( h_{il,jk} + h_{jk,il} - h_{ik,jl} -h_{jl,ik} \ri) \equiv 0~,
\eea
i.e. it vanishes identically.
%
%
That the spatial part of Einstein equations are trivial can also be seen from
\cite{feynman, deser1}.
This is also consistent with the well-known result that the amplitude of gravity waves is related to
the Weyl tensor \cite{york,osano}, which is zero for the FRW Universe.
The overall expansion (preserving homogeneity, isotropy and flatness) is of
course driven by the effective cosmological
constant due to the wavefunction of the graviton condensate, as explained earlier.
Like homogeneity and isotropy, the flatness of space at cosmological scales
is also preserved at every instant of time
including at present, and no fine-tuning in early epochs are needed.
Observed spacetime curvatures and gravity waves at smaller (astrophysical) scales
on the other hand are due to local gravitational effects at the level of stars
and galaxies. Similarly, any gravity waves at cosmological scales can be attributed
to the residual gravitons outside the BEC.

\vs{0.2cm}
\noindent
{\bf Scale invariant spectrum}

The commonly held view is that
the minute quantum fluctuations of the scalar inflaton field and that of the
spacetime metric coupled to it, and the `freezing' of long wavelength modes
outside the horizon during inflation, and their subsequent re-entry at a later time
are responsible for the scale invariant spectrum of perturbations over the uniform background,
and also the observed large scale structures in our Universe \cite{dodelson,weinberg,baumann}.
But as pointed out in \cite{wald}, slow-roll conditions are not essential for the above.
In lieu of the inflaton field, now the wavefunction $\phi$, as well as its fluctuations satisfy
the relativistic Gross-Pitaevskii equation (with self-interaction strength $g$)
$\le[ \Box + m^2 + g |\phi|^2 \ri]\phi =0$
\cite{morikawa,liberati}, or a suitable generalization thereof.
As noted earlier, due to its infinite heat conductivity, the BEC
with uniform temperature throughout is established rapidly, both inside
and outside the horizon radius. Therefore quantum fluctuations also occur almost
simultaneously throughout, and modes outside the horizon are frozen.
Subsequently as the Hubble radius expands,
more and more of such modes enter the horizon.
For a mode $k$ entering the horizon when the
scale factor is $a_B$, and the Hubble parameter $H_B$, such that $k/a_B=H_B$, one has
\bea
\le( \Delta \phi_k \ri)^2 \simeq \frac{1}{2a_B^3(k/a_B)} \simeq \frac{H_B^2}{k^3}~,
\eea
corresponding to a scale invariant spectrum, just as observed in CMBR anisotropies \cite{wald}.
This is also seen from the perturbation equations.
%
The formation of modes throughout almost simultaneously implies the
relation $\eta \simeq -1/a_B H_B$ between conformal time and expansion rate,
leading to the equation for perturbations \cite{dodelson}
\bea
\ddot v + \le(k^2 - \frac{2}{\eta^2} \ri) =0 ~,
\eea
(where $v$ denotes scalar or tensor perturbations, and overdot signifies derivative with respect to
the conformal time)
which again predicts a scale invariant spectrum.
This is consistent with the fact that inflation although sufficient, is not necessary for a scale invariant
spectrum \cite{dodelson}.

\vs{0.2cm}
\noindent
{\bf Conclusions}

Starting with just one reasonable assumption, namely that gravitons have a tiny mass, consistent with
theory and experiments, we have shown that the corresponding critical temperature for condensate formation
is extremely high. Thus they will form a condensate at the earliest epochs.
The density of the BEC accounts for the DM in our Universe, while an effective cosmological constant
derived from the quantum potential of the condensate accounts for the observed DE.
Furthermore a macroscopic wavefunction describing the BEC made up of gravitons in their ground states
is homogeneous and isotropic by nature, and in the absence of propagating gravitons, space is effectively flat
for any time-slice.
No fine tuning is required, and the laws of thermodynamics are satisfied; in fact the latter
require the formation of the condensate.
Our model appears to be at least as effective as inflation in explaining the above observations,
but with far fewer independently unverifiable assumptions. Other problems associated with inflation,
such as reheating, trans-Planckian problem, or that of self-reproduction or multiverses are
avoided here as well.
Further due to the continuous growth of the cosmic BEC,
our model predicts that homogeneity, isotropy and flatness should be preserved
at {\it all} epochs, past and future, unlike inflation, which does not rule out spatial curvatures at
very early and late epochs. This is a prediction which may be used to distinguish it from the inflationary paradigm.
Since the BEC is present forever (unlike the inflaton, which decays soon after the inflationary era),
there should be other testable predictions as well.
For example, a comparison of density profiles generated within a BEC at galactic scales
may be compared with those that are suggested by observations and simulations of DM.
Finally, we note that since the BEC forms right after the BB (or earlier, if there was no BB),
there may have not been sufficient time
for GUT monopoles to form, even if the theories which predict their production are taken to be correct.
Therefore this provides a viable solution of the so-called monopole problem as well.


\vs{.2cm}
\no {\bf Acknowledgment}

\no
I thank A. F. Ali and R. K. Bhaduri for useful discussions.
This work is supported by the Natural Sciences and Engineering
Research Council of Canada.
%




\begin{thebibliography}{99}


\bibitem{perlmutter} S. Perlmutter et al.,
Astrophysical Journal {\bf 517 (2)} (1999) 56586 [arXiv:astro-ph/9812133].

\bibitem{riess} A. G. Riess et al., Astron. J. {\bf 116} (1998) 1009 [arXiv:astro-ph/9805201].

\bibitem{wmap} G. Hinshaw et al, Astrophys. J. Suppl. Ser. {\bf 208:19} (2013) 1 [arXiv:1212.5226].

\bibitem{bao} D. J. Eisenstein et al, Astrophys. J. {\bf 633} (2005) 560 [arXiv:astro-ph/0501171].

\bibitem{guth} A. Guth, Phys. Rev. {\bf D23} (1981) 347.

\bibitem{linde} A. Linde, Phys. Lett. {\bf B108} (1982) 389.

\bibitem{earman} J. Earman, J. Mosterin, Phil. Sc. {\bf 66} (1999) 1-49.

\bibitem{penrose} R. Penrose, Ann. N.Y. Acad. Sci. {\bf 271} (1989) 249-264.

\bibitem{turok} N. Turok, Class. Quant. Grav. {\bf 19} (2002) 3449-3467.

\bibitem{gibbons} G. W. Gibbons, N. Turok, Phys. Rev. {\bf D77} (2008) 063516.

\bibitem{branden1} R. H. Brandnburger in {\it Inflationary Cosmology}, eds. M. Lemoine,
J. Martin, P. Peter (2008).

\bibitem{steinhardt} P. Steinhardt, Sci. Am. (2011) 38-43.

\bibitem{steinhardt2}
A. Ijjasa, P. J. Steinhardt and A. Loeb, Phys.Lett. B723, pp. 261-266, 2013.

\bibitem{steinhardt3}
A. Ijjasa, P. J. Steinhardt and A. Loeb, arXiv:1402.6980, 2014.

\bibitem{planckbicep}
P. A. R. Ade et al, Phys. Rev. Lett. {\bf 114} (2015) 101301, arXiv:1502.00612.

\bibitem{ekpy} J. Khoury, B. A. Ovrut, P. J. Steinhardt, N. Turok,
Phys. Rev. {\bf D64} (2001) 123522.

\bibitem{stringgas} R. H. Brandenburger, Class. Quant. Grav. {\bf 28} (2011) 204005.

\bibitem{mukhanov} V. Mukhanov, arXiv:1409.2335.

\bibitem{sd} S. Das, Phys. Rev. {\bf D89} 084068 (2014) [arXiv:1311.6539].

\bibitem{dasessay} S. Das, 
Int. J. Mod. Phys. {\bf D23} (2014) (1442017) [arXiv:1405.4011].

\bibitem{alidas} A. F. Ali, S. Das,
Phys. Lett. {\bf B741} (2015) 276-279 [arXiv:1404.3093].

\bibitem{dasbhaduri} S. Das, R. K. Bhaduri, 
Class. Quant. Grav. {\bf 32} (2015) 105003
[arXiv:1411.0753].

\bibitem{bohm}
D. Bohm, Phys. Rev. {\bf 85} (1952) 166; D. Bohm, B. J.
Hiley, P. N. Kaloyerou, Phys. Rep. {\bf 144}, No.6 (1987) 321.

\bibitem{zwicky} F. Zwicky,
{\it Cosmic and terrestrial tests for the rest mass of gravitons},
Publications of the Astronomical Society of the Pacific, Vol. 73, No. 434, p.314.

\bibitem{mukhanov2} A. H. Chamseddine, V. Mukhanov, JHEP08(2010) 011 [arXiv:1002.3877].

\bibitem{oda} I. Oda, Mod. Phys. Lett. {\bf A25}(2010) 2411-2421 [arXiv:1003.1437].

\bibitem{derham} C. de Rham, G. Gabadadze, L. Heisenberg, D. Pirtskhalava,
Phys. Rev. {\bf D83} (2011) 103516 [arXiv:1010.1780];
C. de Rham, G. Gabadadze, A. J. Tolley,
Phys. Rev. Lett. {\bf 106} (2011) 231101 [arXiv:1011.1232];

\bibitem{hassan} S. F. Hassan, R. A. Rosen, Phys. Rev. Lett. {\bf 108} (2012) 041101 [arXiv:1106.3344].

\bibitem{mann} J. R. Mureika, R. B. Mann,
Mod. Phys. Lett. {\bf A26}  (2011) 171-181 [arXiv:1005.2214].

\bibitem{majid} S. Majid, arXiv:1401.0673.

\bibitem{hinterbichler} K. Hinterbichler,
Rev. Mod. Phys. {\bf 84} (2012) 671-710 [arXiv:1105.3735].

\bibitem{graviton}
C. M. Will, Phys. Rev. {\bf D57} (1998) 2061;
L. S. Finn, P. J. Sutton Phys. Rev. {\bf D65} (2002) 044022 [arXiv:gr-qc/0109049];
A. S. Goldhaber, M. M. Nieto, Rev. Mod. Phys. {\bf 82} (2010) 939 [arXiv:0809.1003]
E. Berti, J, Gair, A. Sesana, Phys. Rev. {\bf D84} (2011) 101501(R).


\bibitem{sudarshan}
K. P. Sinha, C. Sivaram, E. C. G. Sudarshan,
Found. Phys. {\bf 6}, no.1 (1976) 65;
Found. Phys. {\bf 6}, no.6 (1976) 717.

\bibitem{hu}
W. Hu, R. Barkana, A. Gruzinov,
Phys. Rev. Lett. {\bf 85} (2000) 1158-1161 [arXiv:astro-ph/0003365].

\bibitem{morikawa} M. Morikawa,
22nd Texas Symp. on Rel. Astro. (2004) 1122;
T. Fukuyama, M. Morikawa, Prog. Theo. Phys. {\bf 115}, no. 6 (2006) 1047-1068.

\bibitem{moffat} J. W. Moffat, astro-ph/0602607.

\bibitem{wang} X. Z. Wang, Phys. Rev {\bf D64} (2001) 124009.

\bibitem{boehmer} C. G. Boehmer, T. Harko, JCAP 0706:025,2007 [arXiv:0705.4158];
T. Harko, G. Mocanu, Phys. Rev. {\bf D85} (2012) 084012 [arXiv:1203.2984].

\bibitem{sikivie} P. Sikivie, arXiv:0909.0949.

\bibitem{velten} H. Velten, E. Wamba, Phys. Lett. {\bf B709} (2012) 1-5 [arXiv:1111.2032].

\bibitem{dvali} G. Dvali, C. Gomez, Fortsch. Phys. {\bf 61} (2013) 742-767 [arXiv:1112.3359].

\bibitem{chavanis} P-H. Chavanis, A \& A {\bf 537} (2012) A127 [arXiv:1103.2698].

\bibitem{kain} B. Kain, H. Y. Ling, Phys. Rev. {\bf D85} (2012) 023527 [arXiv:1112.4169].

\bibitem{suarez} A. Su\'arez, V. Robles, T. Matos,
Astroph. and Space Sc. Proc. {\bf 38} Chapter 9 (2013) [arXiv:1302.0903].

\bibitem{ebadi} Z. Ebadi, B. Mirza, H. Mohammadzadeh, JCAP 11(2013)057 [arXiv:1312.0176].

\bibitem{laszlo1} M. Dwornik, Z. Keresztes, L. A. Gergely,
{\it Recent Development in Dark Matter Research},
Eds. N. Kinjo, A. Nakajima, Nova Science Publishers (2014), p.195-219 [arXiv:1312.3715];
arXiv:1406.0388.

\bibitem{bettoni} D. Bettoni, M. Colombo, S. Liberati,
JCAP02(2014)004 [arXiv:1310.3753].

\bibitem{gielen} S. Gielen, arXiv:1404.2944.

\bibitem{schive} H. Schive, T. Chiueh, T. Broadhurst, Nature Physics {\bf 10} (2014) 496
[arXiv:1406.6586].

\bibitem{davidson} S. Davidson, Astrop. Phys.{\bf 65} (2015) 101 [arXiv:1401.1139].


\bibitem{brack} M. Brack, R. K. Bhaduri, {\it Semiclassical Physics}, Westview Press (2003).

\bibitem{rogel} J. Rogel-Salazar, Eur. J. Phys. {\bf 34} (2013) 247 [arXiv:1301.2073];
N. \"Uzar, S. Deniz Han, T. T\"ufekci, E. Aydiner, arXiv: 1203.3352.

\bibitem{feynman} R. P. Feynman, F. B. Morinigo, W. G. Wagner,
{\it Feynman Lectures on Gravitation}, Westview Press (2003), Chaps. 3-5.

\bibitem{deser1} S. Deser, Gen. Rel. Grav. {\bf 1}, no.1 (1970) 9-18;
Class. Quant. Grav. {\bf 4} (1987) L99.

\bibitem{york} J. York, Phys. Rev, Lett. {\bf 28} (1972) 1082.

\bibitem{osano} B. Osano, arXiv:1309.2768.


\bibitem{dodelson} S. Dodelson, {\it Modern Cosmology}, Academic Press (2003), Chap.6.

\bibitem{weinberg} S. Weinberg, {\it Cosmology}, Oxford (2008).

\bibitem{baumann} D. Baumann, {\it TASI Lectures on Inflation}, arXiv:0907.5424.

\bibitem{wald} S. Hollands, R. M. Wald, Gen. Rel. Grav. {\bf 34} (2002) 2043-2055.

\bibitem{liberati}
S. Fagnocchi, S. Finazzi, S. Liberati, M. Kormos, A. Trombettoni,
arXiv: 1001.1044.; \\
%
D. Bettoni, M. Colombo, S. Liberati, JCAP02(2014)004 [arXiv:1310.3753].

\end{thebibliography}
\end{document}